\documentclass{article}

\usepackage{arxiv}

\usepackage[utf8]{inputenc} 
\usepackage[T1]{fontenc}    
\usepackage{hyperref}       
\usepackage{url}            
\usepackage{booktabs}       
\usepackage{amsfonts}       
\usepackage{nicefrac}       
\usepackage{microtype}      
\usepackage{lipsum}

\usepackage{cite}
\usepackage{float} 
\usepackage{amsmath,amssymb,amsfonts}
\usepackage{algorithmic}
\usepackage{graphicx}
\usepackage{textcomp}
\usepackage{multirow}
\usepackage{xcolor}
\usepackage{soul}
\usepackage{color}
\usepackage{caption}
\usepackage{subcaption}
\def\BibTeX{{\rm B\kern-.05em{\sc i\kern-.025em b}\kern-.08em
    T\kern-.1667em\lower.7ex\hbox{E}\kern-.125emX}}

\begin{document}

\title{AI-Powered Interfaces for Extended Reality to Support Remote Maintenance}

\author{Akos Nagy \\
Department of Networks \& Digital Media \\
School of Computer Science \& Maths,\\
ECE\ Kingston University\\
Kingston upon Thames, UK\\
\texttt{A.Nagy@kingston.ac.uk} \\
\And
George Amponis \\
Department of Computer Science \\
International Hellenic University\\
Kavala, Greece \\
\texttt{geaboni@cs.ihu.gr} \\
\And
Konstantinos Kyranou \\
Business Development \& \\
Innovation Department \\
Sidroco Holdings Ltd, \\
Nicosia, Cyprus \\
\texttt{kkyranou@sidroco.com} \\
\And
Thomas Lagkas \\
Department of Computer Science \\
International Hellenic University\\
Kavala, Greece \\
\texttt{tlagkas@cs.ihu.gr}
\And
Alexandros Apostolos Boulogeorgos \\
Department of Electrical and  \\
Computer Engineering,  \\
University of Western Macedonia,\\
Kozani, Greece \\
\texttt{ampoulogeorgos@uowm.gr} \\
\And
Panagiotis Sarigiannidis \\
Department of Electrical and \\
Computer Engineering,  \\
University of Western Macedonia, \\
Kozani, Greece \\
\texttt{psarigiannidis@uowm.gr} \\
\And
Vasileios Argyriou \\
Department of Networks \& Digital Media  \\
School of Computer Science \& Maths, \\
ECE, Kingston University\\
Kingston upon Thames, UK \\
\texttt{Vasileios.Argyriou@kingston.ac.uk}
}

\maketitle

\begin{abstract}
High-end components that conduct complicated tasks automatically are a part of modern industrial systems. However, in order for these parts to function at the desired level, they need to be maintained by qualified experts. Solutions based on Augmented Reality (AR) have been established with the goal of raising production rates and quality while lowering maintenance costs. With the introduction of two unique interaction interfaces based on wearable targets and human face orientation, we are proposing hands-free advanced interactive solutions in this study with the goal of reducing the bias towards certain users. Using traditional devices in real time, a comparison investigation using alternative interaction interfaces is conducted. The suggested solutions are supported by various AI powered methods such as novel gravity-map based motion adjustment that is made possible by predictive deep models that reduce the bias of traditional hand- or finger-based interaction interfaces.
\end{abstract}


\section{Introduction}
Modern production systems are provided with sophisticated machinery. However, the machine tools require maintenance conducted by highly qualified professionals in order to function. Due to the necessity for greater production rates and, thus, decreased lead times, machine breakdowns, i.e. Mean Time To Repair (MTTR), as well as the time required to educate new employees must be minimised as much as feasible. Augmented Reality is a digital technology that enables the viewing of computer-generated 3D information in the Field of View (FOV) of the user, superimposed on the physical environment. In the past ten years, numerous research papers have been published concentrating on the creation of augmented reality-based frameworks for training new people and the provision of remote guidance for Maintenance, Repair, and Operations (MRO). However, the development of AR-based frameworks is a skill- and time-intensive process, and present approaches are focused on the provision of standard AR material that is not tailored to user demands. In contrast, the most recent technological developments have led to the creation of computer vision techniques. These techniques are based on the design of digital networks, also known as Artificial Neural Networks (ANN), which are comparable to the human brain. Additionally, these strategies can be taught to mimic human learning behaviour. Consequently, by applying such methodologies in contemporary industrial settings, applications can learn by doing and, more crucially, self-improve. In an effort to connect the aforementioned technologies, this study discusses the conceptualization, design, and first development of a framework based on the use of Convolution Neural Network (CNN) to generate AR MRO instructions automatically. It is anticipated that the implementation of the suggested framework will eventually aid engineers in lowering the time required for the development of AR instructions and the complexity of such applications. Moreover, by exploiting the capabilities of ANNs, the suggested framework can self-improve over time, hence maintaining a high degree of added value.

\section{STATE OF THE ART}

CNNs originated in the field of computer vision with the purpose of recognising the MNIST (Modified National Institute of Standards and Technology) dataset, which consists of sixty thousand grayscale square images of handwritten numbers. Today, CNN has evolved into popular computer vision technologies, including image recognition, object detection, etc. In the literature \cite{b23}, data-driven approaches are often constituted of three steps: fault-related data collecting, development of Health Indicators (HI), and prediction of time to failure. The performance of these strategies depends greatly on the health markers. Based on a survey of the relevant literature, the health indicator deep learning algorithms can be divided into two groups. The first is CNN, whereas the second is based on Recurrent Neural Networks (RNN). To adapt bearing health indicators, an end-to-end recurrent CNN is suggested. This reduces the weakness of the CNN-based technique and the RNN-based method and further characterizes the pattern of nonlinear bearing deterioration as a roughly linear process over time, while bearings operate under a variety of working situations \cite{b24}. Based on the aforementioned study, it can be stated that Neural Networks in general, and CNNs in particular, are extremely adaptable methodologies that may be used to solve a wide range of engineering challenges. In addition, these research efforts provide new options for enhancing the suggested system architecture in order to expand its capabilities, resulting in a more robust solution with enhanced functionality.

In recent years, deep learning has catalysed exponential growth in artificial intelligence and has been widely implemented in a variety of pattern recognition applications \cite{b1}. In addition, the forecast of machines' Remaining Usable Life (RUL) is becoming increasingly dependent on deep learning. Defect detection, particularly in the field of planetary gearboxes based on vibration data, continues to be a significant challenge for both industry and academics \cite{b2}. In this regard, a dynamic CNN ensemble is presented for fault diagnosis of planetary gearboxes that fuses defect information from multi-level wavelet packet coefficients. A similar study has been conducted, describing an enhanced CNN with larger receptive fields for gearbox problem diagnostics. 

Moving on to another application field of CNN, an overview of CNN-based action recognition and a discussion of action recognition performance on large-scale benchmarks are offered. Initially, the analysis of relevant literature revealed the many obstacles to action recognition \cite{b3, b4, b5}. It then examined the approaches to action identification based on their capacity to meet these issues and deployed publicly accessible datasets including documented challenges. Another review performed an experimental analysis on three datasets under centralised conditions and determined the implications of the experimental results and conclusions, while other papers reviewed action recognition, encompassing a wide range of methods, from handcrafted to deep learning representation.

Recurrent convolutional neural network (RCNN) is introduced as a solution to the constraints of RUL for the prediction of machinery by RUL. In addition, a novel deep learning approach based on multi-scale feature extraction is proposed for predicting the remaining usable life (RUL). The proposed method achieves the direct correlation between the raw data and the ground-RUL without using any prior information. In addition, the suggested method for deep learning consists of three multi-scale blocks and multiple conventional convolutional layers. Next, repeating convolutional layers are created in RCNN to represent the temporal dependencies between the various deterioration phases. Deep CNNs have lately shown outstanding performance in a variety of applications, such as speech recognition and machine vision. The work employs the capabilities of the Gabor filter to extract more discriminating picture features by capturing more image information \cite{b6, b7, b8, b9}. The constructed filter bank accommodates numerous alterations to the basic Gabor filter by altering the Gabor parameters and adjusting the filter's shape. This is achieved by altering the updating rule of CNN in order to examine the contribution of each Gabor parameter to each pixel that is to be classified. 

However, only a small number of studies considered CNN-based methods to be part of deep learning representation. In addition, only a few articles integrate two cutting-edge technologies, such as AR and neural networks, which are frequently used. Specifically, a cellular neural network in a unique FPGA-based Augmented Reality system geared toward assisting visually handicapped persons is analysed. Last but not least, the development of a 3D integral image-based AR with Deep Learning accomplished using faster RCNN at a low-cost lens is detailed. 

Maintenance and its expenses continue to attract the attention of production management over the years since unplanned system breakdowns reduce system reliability and investment returns. Manufacturers should integrate advanced monitoring systems so that machine tool problems can be easily spotted before they occur in order to increase their efficiency through more precise and expedient maintenance. In order to do this, a cloud-based platform for condition-based preventative maintenance is offered as a Product-Service System (PSS), accompanied by a shop floor monitoring service and an augmented reality (AR) application \cite{b10, b11}. Referring to maintenance, the context-aware Augmented Reality System proposed to aid operators in regular maintenance operations by delivering context-relevant information has been researched extensively, and maintenance through AR is provided using a product-service model. In addition, an augmented reality (AR)-based tool is introduced that will improve Business-to-Business (B2B) and internal communication through the usage of smart devices, as well as minimise maintenance time and costs. 

\begin{figure}[h]
\centerline{\includegraphics[width=0.8\textwidth]{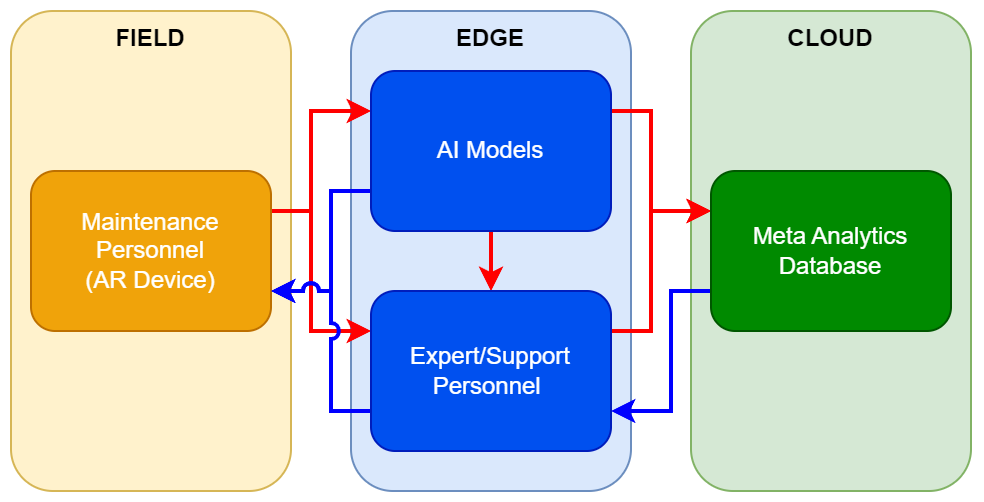}}
\caption{Visualisation of the proposed architecture}
\label{fig:Architecture}
\end{figure}

According to the Industry 4.0 concept, the works \cite{b12, b13, b14, b15} describe an integrated system composed of a machine tool monitoring tool and an AR mobile application that interfaces with a shop floor scheduling tool, as visualised in Fig. \ref{fig:Architecture}. In addition, the estimation of the maintenance time for a new maintenance project, which is one of the most important maintenance offerings, is based purely on the experience and knowledge of the engineer \cite{b16, b17, b18, b19, b20, b21, b22 }. To increase the capture and reuse of expertise in maintenance activities and to enhance the performance of the supplied maintenance PSS, this work presents a knowledge-based maintenance time estimation method based on the analysis of Key Performance Indicators (KPIs).

\begin{figure*}[h]
\centerline{\includegraphics[width=0.9\textwidth]{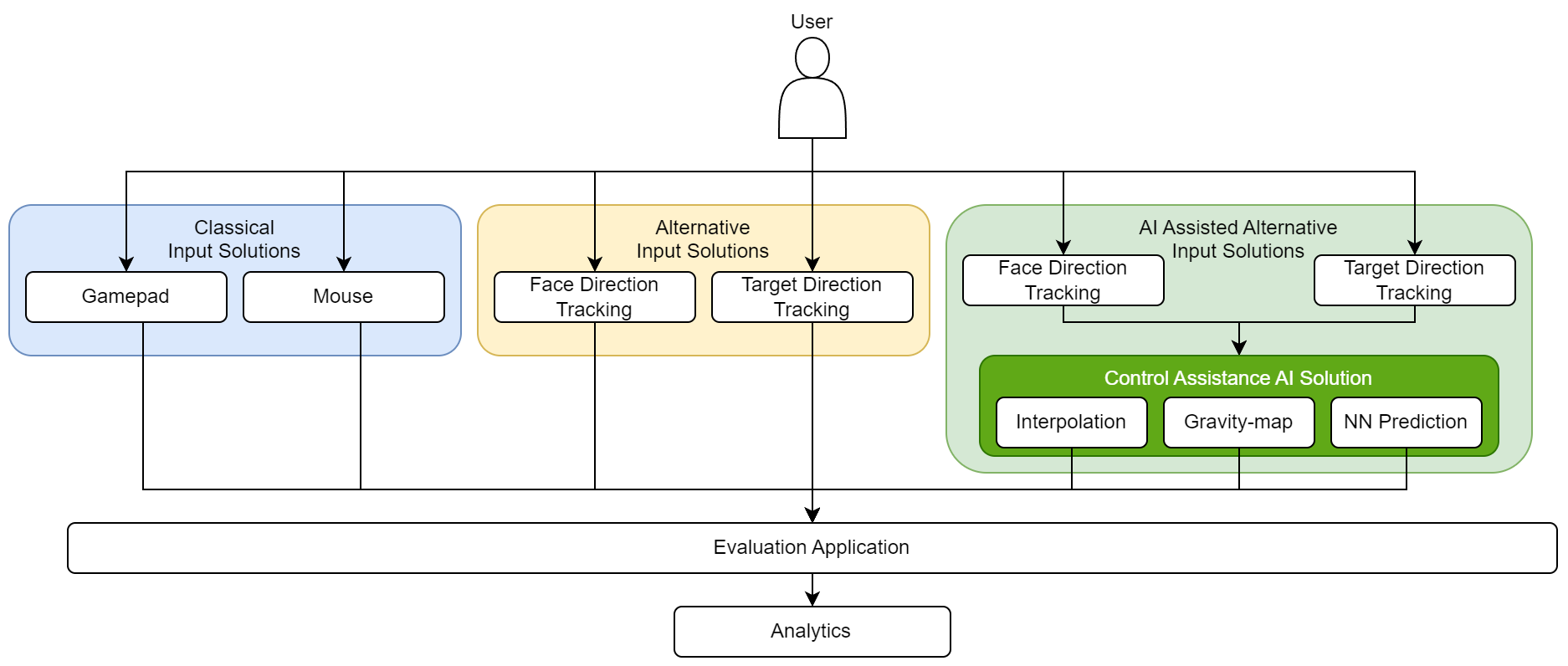}}
\caption{Visualisation of the proposed framework’s interfaces}
\label{fig:Framework}
\end{figure*}

\section{METHODOLOGY}

\subsection{AR Maintenance}

Well-kept industrial equipment is one of the most essential production factors in a manufacturing line. In this paper, an AR solution has been proposed to increase the efficiency of the maintenance procedure and reduce the delay brought on by malfunctioning equipment. There are four interconnected components that make up the architecture. The maintenance staff are outfitted with optical see-through head-mounted displays on the field, where the manufacturing machines are in use (AR glasses). The device's optical see-through feature enables users to inspect their surroundings without major visual obstructions or digital artifacts that can be seen on a digital pass-through head-mounted display, while the digital display's augmentation feature allows for the provision of extra information during maintenance. Currently available AR glasses are equipped a spectrum of sensors, including visual and audio modules, allowing for the recording, processing and eventually transfer of multimedia feed streams via a given network. To achieve this, data streams provided by the AR device are relayed to two other architectural components of the proposed synthesis. The aforementioned interfaces are visualised in Fig. \ref{fig:Framework}.
A dedicated device on the edge hosts one or more AI models that can take in and process the data. The outcomes of the AI models, such as the ability to recognize objects and targets, are fed back to the maintenance staff and shown on the AR gadget. The models can be utilised with computer vision techniques to identify equipment, components, and faults.
The expert staff is connected to the architecture via a network accessible device and forms the second element at the edge. Expert employees on the field who are receiving the multimedia feed from the AR gadget may assess the problem almost immediately and help the maintenance workers. When communicating with the expert staff, the AR device can transmit text messages or audio information (such as object highlights) about objects. The outcomes of the AI-powered processes allowing a larger variety of information to reduce the maintenance procedures can also be advantageous to expert employees. Due to the nature of edge devices, the specialist staff may be located on- or off-site, which could allow manufacturers to assist manufacturing facilities in the event of a technical problem.
The Cloud-based Meta Analytics Database is the final element of the architecture. Based on the data provided by the field maintenance workers, the AI models on the Edge extract meta data. The professional staff can then access the processed and saved meta data, offering a long-term overview of operating and maintenance features. The status of the machines, probable abnormalities, and impending maintenance needs can all be predicted using this data.

\subsection{AR Maintenance}

Some individuals may experience difficulties utilising gadgets with traditional input methods due to physical restrictions. Said restrictions may be brought on by a physical impairment or injury in any situation, including industrial settings where the user has to utilise both hands for the operation. The mouse and gamepad are regarded as traditional input devices in the current architecture; the former is frequently used with personal computers, while the latter is the primary input device for console gaming. The project's objective is to define a generalized architecture and solutions that enable the efficient use of a range of alternative input devices in addition to the traditional input device. These suggested remedies and ideas are analysed using both qualitative (e.g., MoS) and quantitative (duration, success rate, accuracy) metrics in a comparison study on a set of trials that comprise object clicking, selection, and follow.
The project's purpose is to develop a generic architecture and solutions that enable a number of alternative input devices to be utilised with the same efficiency as traditional input devices. In the context of a comparative research, these proposed solutions and concepts are evaluated using both qualitative (e.g., MoS) and quantitative (time, success rate, accuracy) criteria.
We are examining alternate, AR-based input options for the aforementioned comparative study and methods to improve the effectiveness of these solutions including AI support in circumstances when traditional input devices are not suitable. Two AR-based input approaches have been considered. The first one is a face tracking solution leveraging a computer webcam, detecting the user's face and direction thereof and providing the underlying components with the relevant input. The second one is a target tracking solution; said solution receives a target image or object position and direction as inputs, and uses this data to control the output accordingly.
The conducted comparative analysis demonstrates the disparity between the traditional input devices and the recommended alternative options. Several AI-based strategies have been developed to reduce these discrepancies. The first solution, depending on the movement direction supplied by the input device, employs linear interpolation. It deviates the input movement in the direction of the specified target location. The deviation considers the distance from the target location as well as the input movement vector in order to give reliable help, while avoiding a forced experience for the user. This method only allows for one target location at a time.
The input movement vector is deviated by the gravity-map solution based on a list of several target coordinates according to the distance of the pointer from the targets. As the user approaches the objectives, the divergence gets bigger. This method enables a dual area design, where certain portions of the application are not impacted by the AI aid, providing a multi-target assistance system that outside of the target regions will not affect the user input. The third option estimates the best movement vector using a machine learning model based on neural networks. The solution shifts the user input in the direction of a target based on the past N locations of the cursor during a specified time period, sampled at predetermined time intervals. A set-sized grid that shows the locations of prospective targets as additional input for the model enables multi-target help.

\section{Results}

\begin{figure}[h]
\centerline{\includegraphics[width=0.45\textwidth]{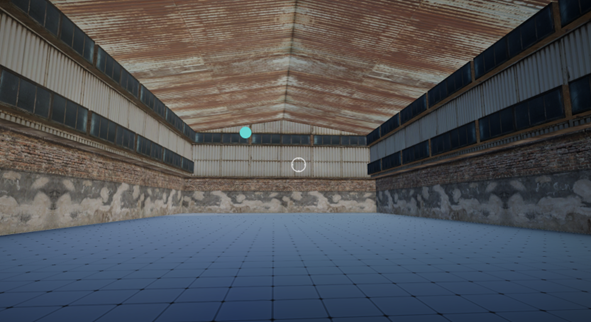}}
\caption{Sample image of the evaluation application featuring a single static target (cyan)}
\label{fig:AppSampleLocate}
\end{figure}

Initial tests were conducted to establish a baseline for the most effective and user-friendly solution, while gathering comparative data on the performance of the various input devices and solutions. Applying comparison research, the input techniques were analysed using both qualitative (such as MoS) and quantitative (duration, success rate, accuracy) metrics on a series of trials involving object clicking, selection, and follow. Fig. \ref{fig:AppSampleLocate} illustrates the evaluation application from which the presented results stem.

\subsection{Locate Mode}

In this mode, a number of static targets are shown in a 3D setting. The participant's objective is to position the crosshair, which represents the screen's center, above the target. Each job consists of a number of targets, each of which is a separate subtask. The targets are organised in a sequential manner, with only one target being available at any given moment. We can determine the average amount of time it takes a participant to get at the goal position by running the experiment with various input methods and solutions. Every target has a predetermined window of availability beyond which the target is deleted, and the assigned subtask is deemed to have failed.

\begin{table}[h]
\centering
\caption{OVERALL OUTCOMES OF LOCATE MODE}.
\label{table:LocateMode}

\begin{tabular}{|l|c|c|c|c|}
    \hline
    & Head & Image & Mouse & Controller \\ \hline   
    Success & 92.5\% & 90.3\% & 98.3\% & 97.8\% \\ \hline
    Min. Screen Dist. & 10px & 14px & 5px & 8px \\ \hline
    Max. Screen Dist. & 1092px & 954px & 1089px & 1009px \\ \hline
    Avg. Screen Dist & 252px & 284px & 300px & 269px \\ \hline
    Avg. Score. & 1.84 s & 2.04 s & 0.89 s & 1.34 s \\ \hline    
    Avg. Distance & 30.97m* & 30.04m* & 30.94m* & 30.89m* \\ \hline    
\end{tabular}\\
\textit{*Scale of distance for test environment to real environment is 1 unit : 1 m}
\end{table}

\begin{figure}[H]
\centering
\begin{minipage}{.45\textwidth}
\centering
\captionsetup{width=.8\linewidth}
\includegraphics[width=0.9\textwidth]{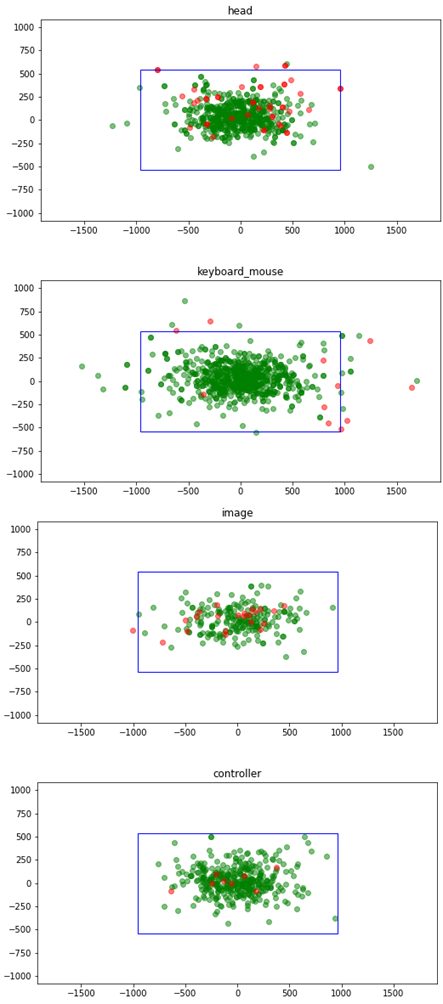}
\caption{Target position on screen the moment of appearing; (green – successful subtask, red – failed subtask)}
\label{fig:LocateModeScreen}
\end{minipage}%
\begin{minipage}{.45\textwidth}
\centering
\captionsetup{width=.8\linewidth}
\includegraphics[width=0.9\textwidth]{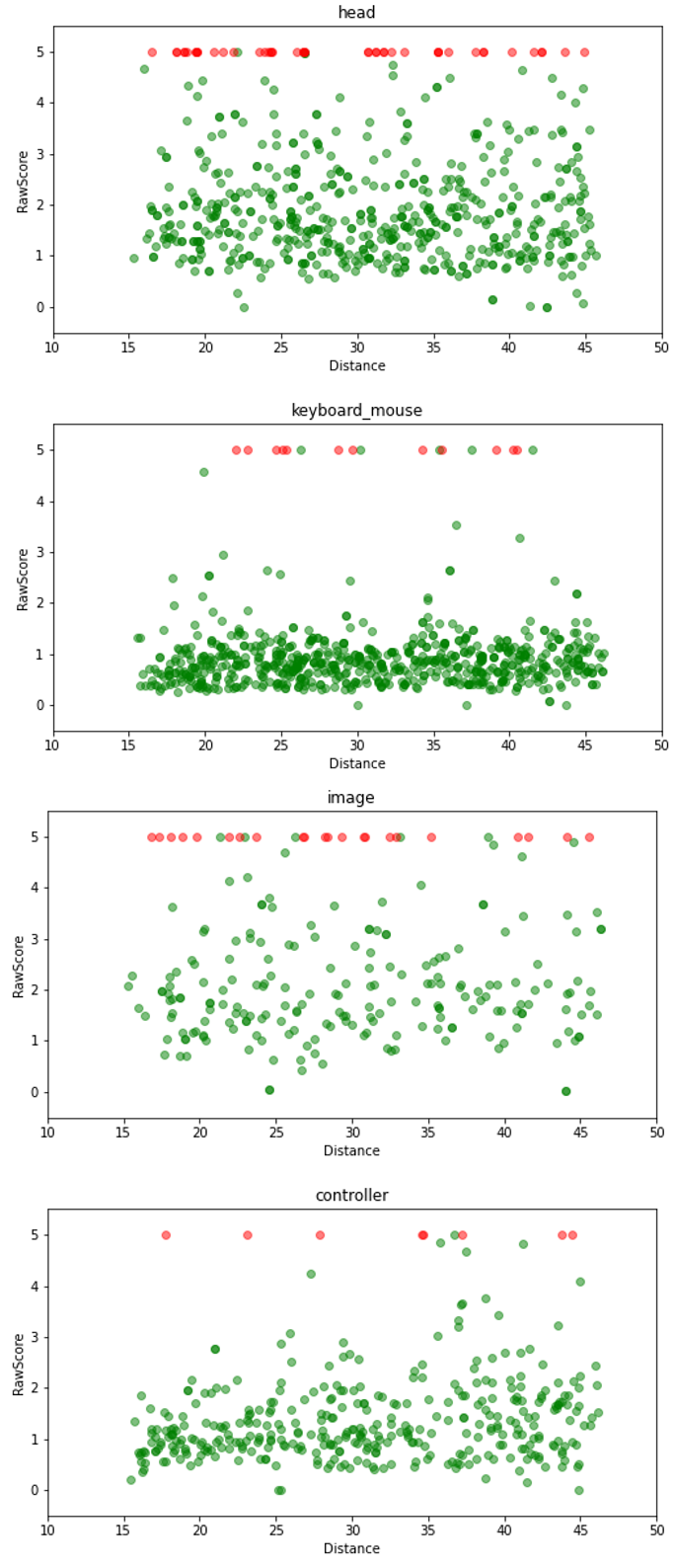}
\caption{Distance of targets from the camera in relation with the additional time it takes to complete the subtask (green – successful subtask, red – failed subtask)}
\label{fig:LocateModeDistance}
\end{minipage}
\end{figure}

According to the preliminary data analysis, which excluded input augmentation, the traditional input devices marginally outperformed the other methods in terms of success rate. If the subject was able to move the crosshair onto the target item in this experiment, the subtask was successfully completed. Based on the successfully completed subtasks, the traditional input methods are typically 30\% to 45\% faster than the other options. The mouse, the most popular input method, performed best overall while the image tracking system fared the poorest.
We also examined how these atypical circumstances would affect the success rate when the target might appear outside of the camera's field of vision in the application. Only the mouse has an observable quantity of failed subtask targets close to the edge of the screen, and the data does not clearly suggest these circumstances as the major contributors of the failed subtasks for the majority of input solutions. This would suggest that the best control strategy should have only failed subtasks when the participant could not see the target item. Table I summarises the results obtained in tests conducted, while Fig. \ref{fig:LocateModeScreen} and \ref{fig:LocateModeDistance} illustrate the success of tasks corresponding to targets Fig. \ref{fig:LocateModeScreen} shows the correlation between target appearance on the screen and task success, while Fig. \ref{fig:LocateModeDistance} shows the correlation between target distance from the camera and task success.

\begin{figure}[h]
\centerline{\includegraphics[width=0.5\textwidth]{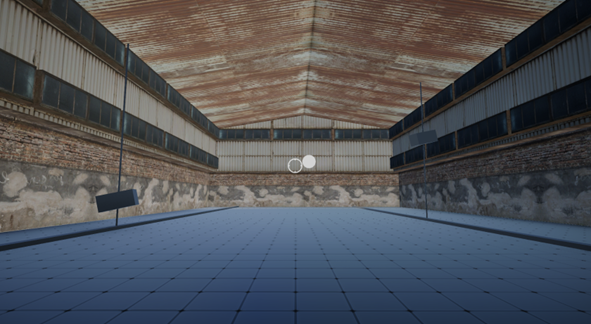}}
\caption{Sample image of the evaluation application featuring a single 
dynamic target (gray) moving between the two path indicators (Follow Mode)}
\label{fig:AppSampleFollow}
\end{figure}

Because the targets appear in a 3D world, the influence of the target's distance from the camera and, in turn, the target's relative size, are thought to be decisive factors in terms of success rate. Based on the original data, neither the success rate nor the time required to complete the supplied subtasks successfully showed any appreciable difference. Fig. \ref{fig:AppSampleFollow} visualises the evaluation environment.

\subsection{Select Mode}

\begin{figure}[H]
\centering
\begin{minipage}{.45\textwidth}
\centering
\captionsetup{width=.8\linewidth}
\includegraphics[width=0.9\linewidth]{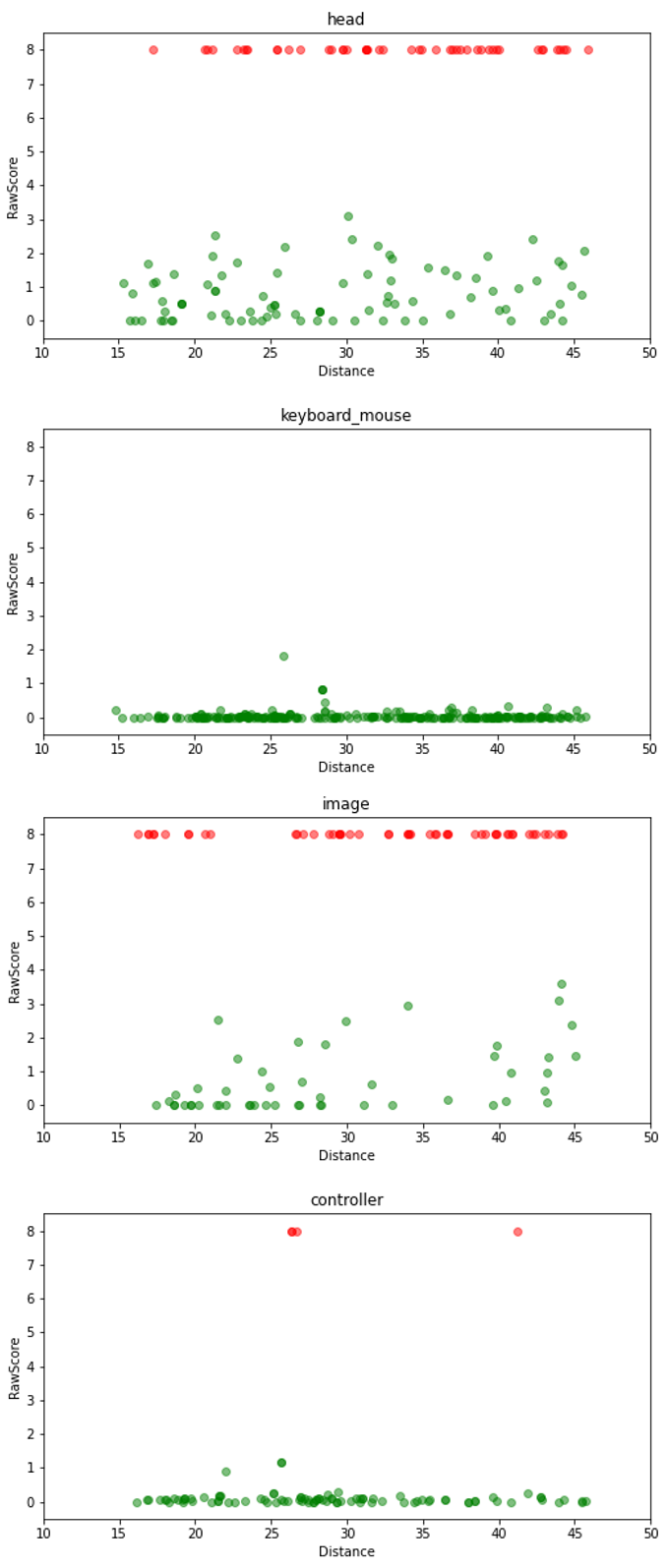}
\caption{Distance of targets from the camera in relation with the time it takes to complete the subtask (green – successful subtask, red – failed subtask)}%
\label{fig:SelectModeDistance}%
\end{minipage}%
\begin{minipage}{.45\textwidth}
\centering
\captionsetup{width=.8\linewidth}
\includegraphics[width=0.9\textwidth]{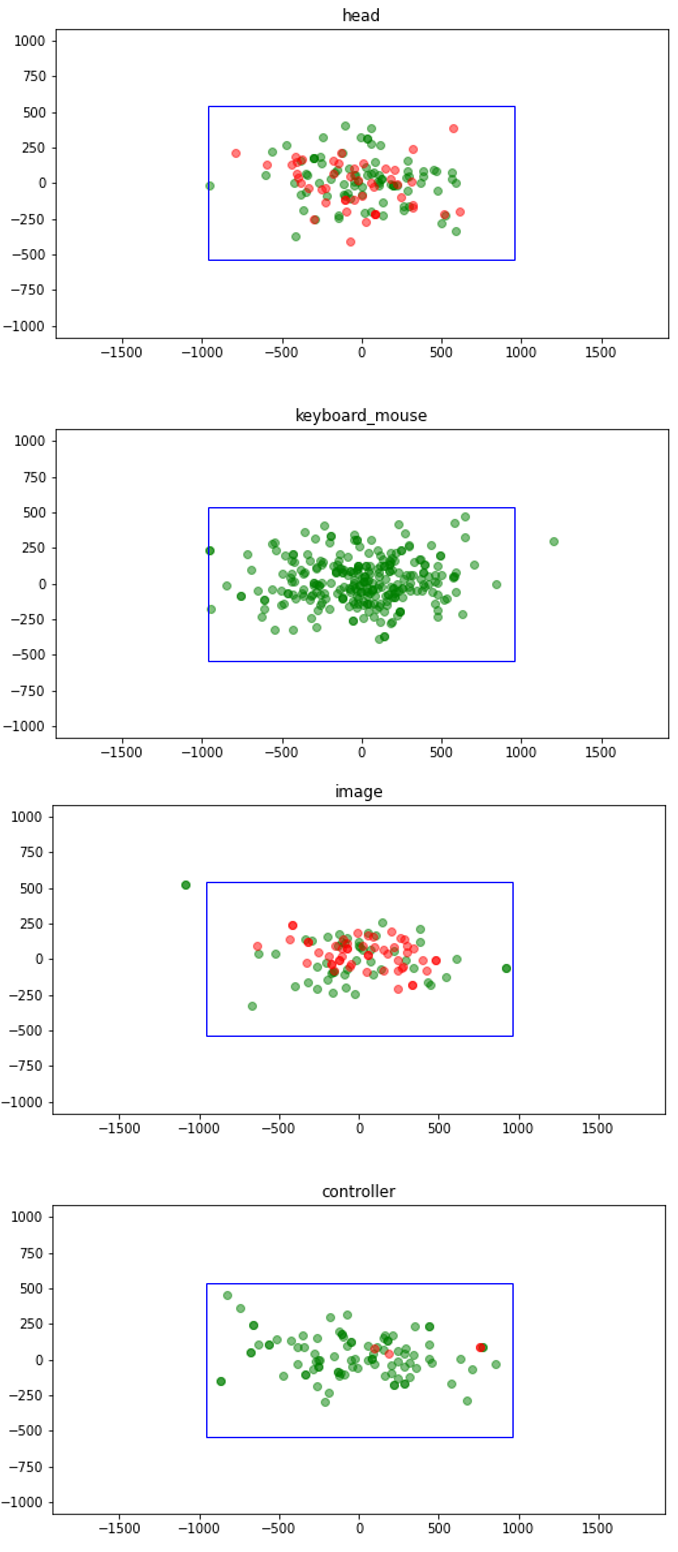}
\caption{Target position on screen the moment of appearing; green successful subtask, red – failed subtask;}%
\label{fig:SelectModeScreen}%
\end{minipage}
\end{figure}

This mode includes a collection of static targets in a 3D environment, similarly to what the Locate Mode does. The participant's objective is to move the crosshair, which serves as the screen's center, onto the target and maintain it there continuously for a certain amount of time. Each job consists of a number of targets, each of which is a separate subtask. The targets are organised in a sequential manner, with only one target being available at any given moment. We may assess the accuracy of an input device by running the experiment with various input methods and solutions. By accuracy, we mean the ability of the input method to stay inside a specific region of the screen.
Every target has a predetermined window of availability beyond which the target is deleted and the assigned subtask is deemed to have failed. When a subtask is completed, the experiment calculates the extra time the crosshair spent on the target during the subtasks that took longer than necessary to complete. Preliminary data collection demonstrates that the traditional input devices greatly outperform the alternative alternatives in terms of success rate, ranging from 45\% to 100\%. On average, other solutions might take up to 19 times longer on the target than the mouse, the input that performs the best. This shows that the other solutions' accuracy is noticeably less stable. Table II and Figs. \ref{fig:SelectModeScreen} and \ref{fig:SelectModeDistance} illustrate the results obtained in this test.
\begin{table}[h]
\centering
\caption{OVERALL OUTCOMES OF SELECT MODE}.
\label{table:SelectMode}
\begin{tabular}{|l|c|c|c|c|}
    \hline
    & Head & Image & Mouse & Controller \\ \hline 
    Success & 65.4\% & 49.0\% & 100\% & 95.8\% \\ \hline
    Min. Screen Dist. & 24px & 18px & 13px & 31px \\ \hline
    Max. Screen Dist. & 951px & 920px & 977px & 941px \\ \hline
    Avg. Screen Dist & 292px & 298px & 307px & 351px  \\ \hline
    Avg. Score. & 0.793 s & 0.712 s & 0.042 s & 0.099 s \\ \hline
    Avg. Distance & 28.97m* & 29.19m* & 31.12m* & 28.76m* \\ \hline
\end{tabular}\\
\textit{*Scale of distance for test environment to real environment is 1 unit : 1 m}
\end{table}

The effect of the targets appearing outside of the participant's field of view was taken into account as a determining element of success, similar to the Select Mode. The success of the subtasks does not appear to be correlated with how the target appears, according to the preliminary data. The participants could plainly see all of the objectives that appeared in the unsuccessful subtasks. 

The size of the object has a tangential, but not quite conclusive relationship to how distance affects it. In the instance of the image-based input solution, there is a clear indication of where the object's distance increased and the additional time it took to complete the assignment, showing that the obviously smaller items are more challenging to choose appropriately. Despite the technological link between both solutions, this pattern was anticipated but not readily seen on the head input solution. 

\subsection{Follow Mode}

In the same way that the other modes do, the follow mode has a number of subtasks, but in this task, the target is a dynamic object rather than a static one. Between a starting position and an ending point, the targets are moving steadily. To regulate the experiment and determine the relationship between success and speed, we established a set of predefined speeds. Move the crosshair, which represents the screen's center, onto the target object with the intention of following it along the route between the start and end marks. This assignment assesses the precision of the input solutions in a dynamic target scenario, similarly to the Select Mode.

\begin{table}[h]
\centering
\caption{OVERALL OUTCOMES OF FOLLOW MODE}.
\label{table:FollowMode}
\begin{tabular}{|l|c|c|c|c|c|}
\hline
Speed & Label & Head & Image & Mouse & Controller \\ \hline   
\multirow{2}{4em}{2m/s*} & Success & 100.0\% & 100.0\% & 100.0\% & 100.0\% \\ \cline{2-6}
& Score & 26.9\% & 23.5\% & 73.7\% & 32.7\% \\ 
\hline
\multirow{2}{4em}{5m/s*} & Success & 80.7\% & 92.3\% & 98.3\% & 100.0\% \\ \cline{2-6}
& Score & 11.8\% & 13.3\% & 46.9\% & 18.8\% \\ \hline
\multirow{2}{4em}{10m/s*} & Success & 46.6\% & 53.8\% & 100.0\% & 97.6\% \\ \cline{2-6}
& Score & 4.9\% & 3.3\% & 21.9\% & 7.5\% \\ \hline
\multirow{2}{4em}{Average} & Success & 74.7\% & 82.9\% & 99.4\% & 99.1\% \\ \cline{2-6}
& Score & 16.8\% & 15.7\% & 48.1\% & 19.3\% \\ \hline    
\end{tabular}\\
\textit{*Scale of distance for test environment to real environment is 1 unit : 1 m}
\end{table}

If the participant is able to move the crosshair onto the target item in Follow Mode, the subtask is judged successful. The software calculates how long the crosshair is over the target item and uses that information to give a score that represents how much of the total trip time has been spent on the object. There were several different speed categories established for the original trial. All input alternatives were equally successful at lower speeds in terms of success, however as speed increases, we observe a 20\%–30\% increase in success when utilising traditional input devices. The discrepancy is more obvious when we consider the average score. 

Although the controller has a better score, it only significantly exceeds the alternative input options by 15\% to 25\%, but the mouse has a score that is x2 to x5 times that of the alternatives, making it the clear winner when working with dynamic targets.
The graphs and output data clearly show the relationship between the object's speed, where the "FlyTime" is larger, and the likelihood that the subtask would be successful based on the original data. Within the same speed category, the variations in trip lengths and their relationship to screen time do not significantly alter the success rate.

\begin{figure}[H]
\centering
\begin{minipage}{.45\textwidth}
\centering
\captionsetup{width=.8\linewidth}
\includegraphics[width=0.9\textwidth]{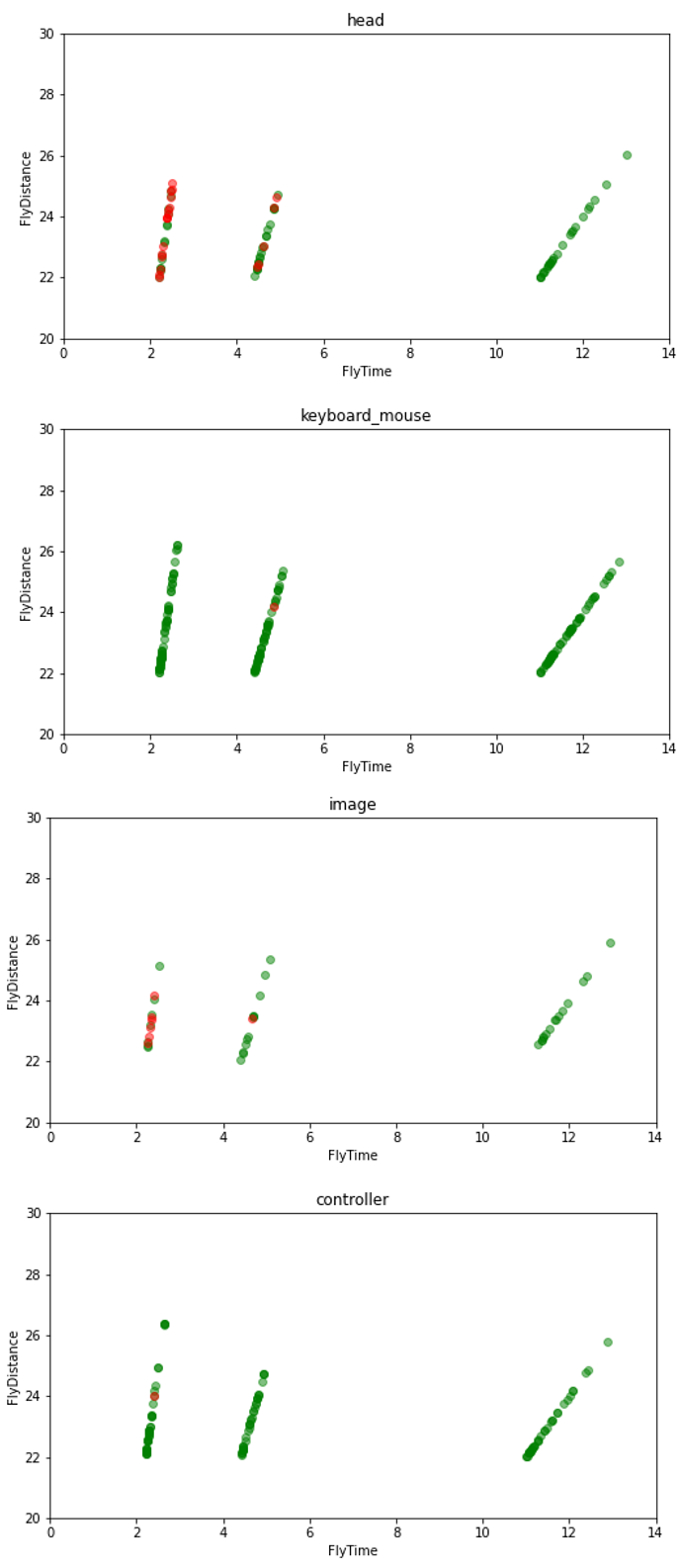}
\caption{Subtask success in relation with the FlyTime and FlyDistance; green – successful subtask, red – failed subtask}
\label{fig:FollowModeDistanceFlytime}
\end{minipage}%
\begin{minipage}{.45\textwidth}
\centering
\captionsetup{width=.8\linewidth}
\includegraphics[width=0.9\textwidth]{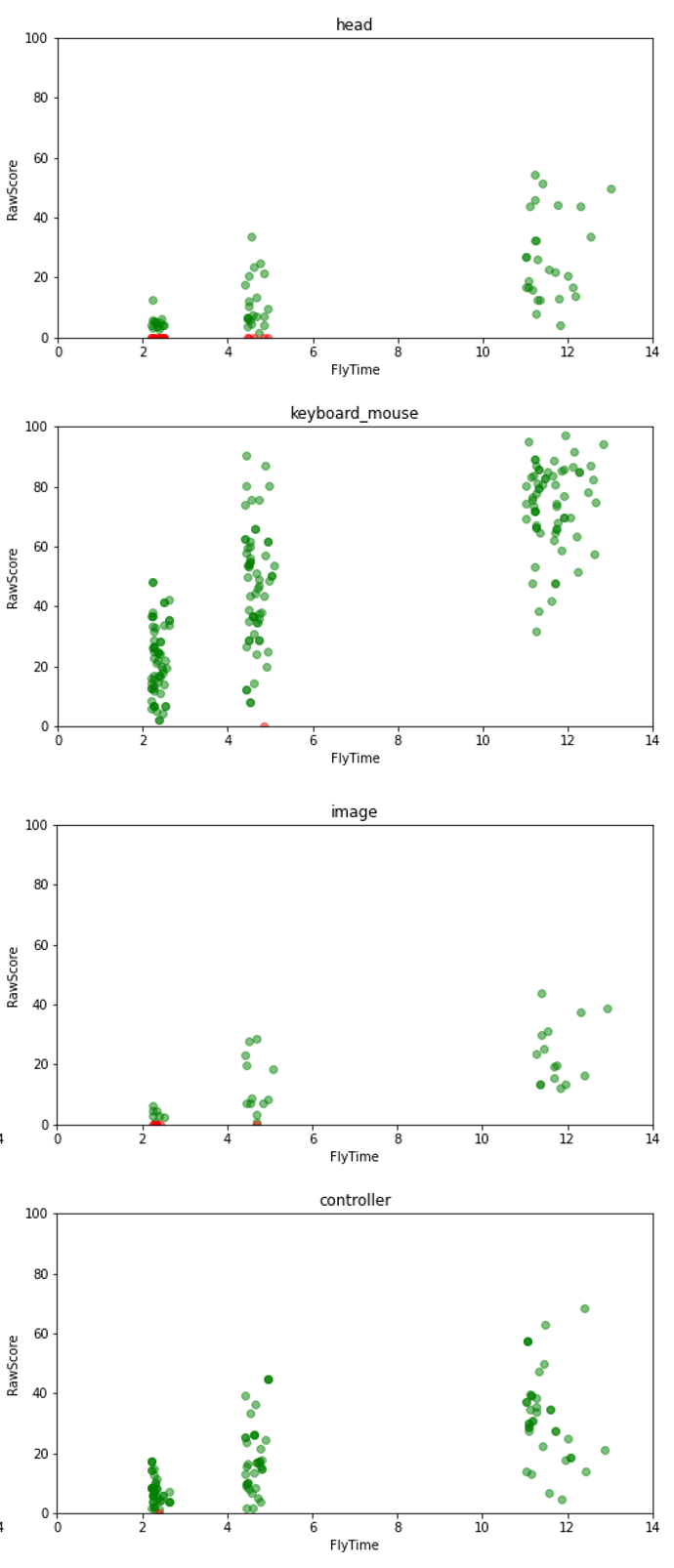}
\caption{The follow percentage in relationship with the FlyTime; green – successful subtask, red – failed subtask}
\label{fig:FollowModeScoreFlytime}
\end{minipage}
\end{figure}

Similarly to the success rate, the score has a clear relationship with the speed category, as illustrated in the chart below. It is abundantly evident that the speed of the dynamic object in the 3D environment is the primary determinant of the rate of the following capabilities. Table III and Figs.  \ref{fig:FollowModeDistanceFlytime} and \ref{fig:FollowModeScoreFlytime} illustrate the results obtained throughout tests in Follow mode.

\section{CONCLUSIONS}

Many Human-Computer Interaction interfaces for immersive environments have been utilised to enable AR technology, allowing more natural and intuitive communication. Such systems are crucial for situations when maintenance workers must use their hands to do chores while interacting with an augmented environment, as well as for elderly and disabled users for whom traditional interfaces may be inaccessible owing to physical limitations. This paper presents a generalised framework and architecture that enables the effective use of a range of non-biased interfaces for maintenance applications or by individuals with special requirements. Applying comparative research, these suggested solutions and concepts have been tested using both qualitative (such as MoS) and quantitative (such as time, success rate, and accuracy) metrics on a set of tasks that include object clicking, selection, and follow.

\section{FUTURE WORKS}

As the preliminary data has been collected and evaluated the research steps into the testing phase. This portion of the research process involves the addition of AI assistance solutions to the testing application. The three methods, as discussed : linear interpolation, gravity-map, and machine learning with neural networks. These methods are designed to assist the user in controlling interfaces using AR technologies and minimize the discrepancy between the classical input solutions and the alternative input methods.

\section{ACKNOWLEDGMENT}

This project has received funding from the European Union’s Horizon Europe Research and Innovation programme under grant agreement No. 101070181.

\end{document}